\documentstyle[a4wide,12pt]{article}

\title{Wave function of the radion with a bulk scalar field}
\author{Ph.~Brax \\
{\it Service de Physique Th\'eorique}\\
{\it CEA-Saclay}\\
{\it F-91191, Gif/Yvette cedex, France}\\
{\it }\\
C.~van~de~Bruck, A.~C.~Davis, C.~S.~Rhodes \\
{\it Department of Applied Mathematics and Theoretical Physics}\\
{\it Centre for Mathematical Sciences}\\
{\it Wilberforce Road, University of Cambridge}\\
{\it CB3 0WA, United Kingdom}}
\begin{document}
\maketitle

\abstract{
The behaviour of the distance between two branes (the `radion') in a
braneworld model with a bulk scalar field is investigated. We show 
that the BPS conditions of supergravity ensure that the dynamics of the 
scalar field and the radion are not independent; we derive the
four-dimensional effective action, showing that the effective theory
is of scalar--tensor nature, coupling the radion to four-dimensional
gravity.}

\vspace{0.5cm}
\noindent DAMTP--2002--002, t02/008
\vspace{0.5cm}
\section{Introduction}
The idea that we live on a hypersurface embedded in a
higher--dimensional space sparked a lot of interest 
amongst  cosmologists and particle physicists. It is very conceivable
that the brane--world idea sheds light 
on some unsolved problems of our standard models of the 
micro-- and macro--worlds. The idea is mainly motivated by
recent progress in string-- and M--theory \cite{Lukas}, but 
phenomenological approaches led to important insights. In particular, 
the model by Randall and Sundrum plays the r{\^o}le of a simple 
playground where the brane world idea can be tested 
\cite{Randall1},\cite{Randall2}. However, 
as with all simple models, it might not be realized in nature. 
One important extension is to consider matter fields in the 
bulk, such as scalar fields (see \cite{BD1}--\cite{Flanagan} and
references therein). These matter fields exist in some fundamental
approaches to brane worlds, such as heterotic M--theory \cite{Lukas}
and supersymmetry \cite{Kallosh}, and could affect the
brane matter fields, because of the coupling between them.

In this paper we discuss the dynamics of a brane-world 
scenario with two branes in the presence of a bulk scalar field. 
The second brane is separated from the first brane by a distance 
$r_c$. From the viewpoint of the four--dimensional theory the 
distance between the branes is a scalar degree of freedom, called 
the radion. The dynamics of the radion has been discussed 
in many papers (see \cite{Charmousis}--\cite{radion-x} and references
therein). In this paper we study in particular the coupling 
of the radion to the bulk scalar field. To do so we focus on 
linear perturbations around a static solution. In particular 
this involves solving the equations of motion for the scalar 
and gravitational perturbations. As usual we isolate two 
types of gravitational degrees of freedom: the first type comprises 
the gravitons and its imprints on the brane; the second type springs 
from the existence of two branes and leads to the radion modulus 
parameterizing the proper distance between the branes. In the case of 
the models considered here, the coupling between the bulk scalar field
and the radion ensures that there is essentially only one scalar 
degree of freedom in the four--dimensional effective theory. 

The paper is organized as follows: In section 2 we present the set--up 
of the theory and the corresponding graviton wave function. In section 3 
we discuss the fluctuation of the distances between the branes and 
its coupling to the bulk scalar field. The effective low energy scalar-tensor 
theory is presented.
We summarize our findings in section 4.

\section{The Graviton and the Scalar Field Zero Mode} 
In this section we describe the brane world setup and describe the 
graviton and scalar field zero modes. 
\subsection{ The Background Configurations}
We consider a bulk Lagrangian consisting of two terms which describe 
gravity and the bulk scalar field, respectively:
\begin{equation}
S_{\rm bulk}=\frac{1}{2\kappa_5^2}\int d^5 x
\sqrt{-g_5}(R-\frac{3}{4}((\partial \phi)^2 +U))
\end{equation}
The boundary terms read
\begin{equation}
S_{\rm brane}=-\frac{3}{2\kappa_5^2}\int d^5 x\sqrt {-g_5}(\delta
(z)-\delta (z-r_c))U_B
\end{equation}
where $U_B$ is the superpotential related to the potential as
\begin{equation}
U=\left(\frac{\partial U_B}{\partial \phi}\right)^2-U_B^2.
\end{equation}
We will also include the Gibbons-Hawking term
\begin{equation}
S_{\rm GH}=\frac{1}{\kappa_5^2}\int_{\delta} d^4x \sqrt{-g^B}K
\end{equation}
where $\delta=\delta(z)-\delta(z-r_c)$ is the boundary of
space-time while $g^B$ is the boundary metric and $K$ the
extrinsic curvature. At the position of the branes we 
impose a $Z_2$--symmetry.

In this paper we investigate solutions which are obtained from 
BPS--like equations, which are given by \cite{BD1,BD2} 
\begin{equation}\label{BPS}
\frac{a'}{a}=-\frac{U_B}{4},\ \phi'=\frac{\partial U_B}{\partial \phi}
\end{equation}
where $'=d/dz$ for a metric of the form
\begin{equation}\label{background}
ds^2 = dz^2 + a^2(z)\eta_{\mu\nu}dx^\mu dx^\nu.
\end{equation}
We will particularly focus on the case where
\begin{equation}
U_B=4k e^{\alpha \phi},
\end{equation}
with $\alpha =1/\sqrt 3,-1/\sqrt {12}$ which was obtained in a theory
with supergravity in singular spaces \cite{BD1}. In that case the 
solution reads
\begin{equation}
a(z)=(1-4k\alpha^2z)^{\frac{1}{4\alpha^2}},
\end{equation}
whereas the scalar field solution is
\begin{equation}
\phi = -\frac{1}{\alpha}\ln\left(1-4k\alpha^2z\right).
\end{equation}
In the $\alpha\to 0$ we retrieve the AdS profile
\begin{equation}
a(z)=e^{-kz}.
\end{equation}
Notice that in that case the scalar field decouples altogether.
In this paper we will assume that the conditions (\ref{BPS}) are 
valid everywhere. 

\subsection{Absence of the Scalar Field Zero Mode}
We now consider perturbations around the background metric
(\ref{background}). In order to understand the origin of the 
radion mode, it was shown in \cite{Charmousis} that it is essential 
to analyse the possible gauge choices which preserve the Gaussian 
normal coordinates (see also section 3 below). In this subsection
we concentrate on the bulk scalar field perturbation and its 
transformation under infinitesimal  coordinate transformations which 
preserve the Gaussian normal coordinate system. 

Consider a brane located at
\begin{equation}
z = f(x^{\mu})
\end{equation}
where the bulk metric has the form (\ref{background}), i.e.\ it reads
\begin{equation}
ds^2_0=dz^2 + a^2(z)\eta_{\mu\nu}dx^{\mu}dx^{\nu}.
\end{equation}
We consider perturbations of this background metric for which the 
perturbed metric has the form of a Gaussian normal coordinate 
system, i.e.
\begin{equation}
\delta g_{zz} = 0 = \delta g_{\mu z}.
\end{equation}
When perturbing this background metric, one can change coordinates 
while preserving the Gaussian normal coordinate system and bringing
the brane back to the origin of coordinates. 
This is achieved by considering vector fields
$\xi$ such that $ z\to z+ \xi^z,\ x^{\mu}\to x^{\mu}+\xi^{\mu}$
which satisfy
\begin{equation}
\xi^z\equiv f(x^{\mu}),\ \xi^{\mu}
=-\int \frac{dz}{a^2}\eta^{\mu\nu}\partial_{\nu}\xi_z
 +\eta^{\mu\nu}\epsilon_{\nu}(x^{\mu}).
\end{equation}
Such a change of coordinates shifts the brane position to
\begin{equation}
z = 0.
\end{equation}
We still have five gauge degrees of freedom, specified by 
$\xi^z$ and $\epsilon^\mu$. 
In such a coordinate system the metric reads
\begin{equation}
ds^2=ds^2_0-\frac{U_B}{2}a^2(z)\xi_zdx^{\mu}dx_{\mu}
-2a^2\left(\int\frac{dz}{a^2}\right)\partial_{\mu}\partial_{\nu}\xi_z
dx^{\mu}dx^{\nu}
\end{equation}
where we have put $\epsilon_{\mu}=0$.
Such a change of metric corresponds to  gravitational modes
which lead to the existence of a radion mode.

Let us now consider the case of a perturbed four dimensional part of
the metric 
\begin{equation}
ds^2= dz^2 + a^2(z)dx^{\mu}dx_{\mu}+ h_{\mu\nu}dx^{\mu}dx^{\nu}
\end{equation}
when  the   brane is still located at $z=\xi_z$.
In the Gaussian system at the origin we can write down 
the boundary conditions using
\begin{equation}
K_{\mu\nu}=\frac{1}{2}\partial_z g^B_{\mu\nu}
\end{equation}
where $g^B$ is the brane metric.
The Israel junction conditions can be expressed as the
identity
\begin{equation}
K_{\mu\nu}-g^B_{\mu\nu}K=\frac{1}{2}T^B_{\mu\nu}
\end{equation}
evaluated on the brane. We have denoted by $T^B$ the energy-momentum
tensor on the brane.

In the coordinate system where the brane is at the origin, the
perturbed metric reads
\begin{equation}\label{splitting}
\hat h_{\mu\nu}=
h_{\mu\nu}- \frac{U_B}{2}a^2(z)\xi_z\eta_{\mu\nu} 
- 2a^2\left(\int \frac{dz}{a^2}\right)\partial_{\mu}\partial_{\nu}\xi_z,
\end{equation}
while the perturbation of the scalar field is
\begin{equation}
\delta \hat \phi=\delta \phi + \frac{\partial U_B}{\partial \phi} \xi_z,
\end{equation}
where $\delta\phi$ is the intrinsic part of the scalar field fluctuation
as measured when the brane is at $z=\xi_z$. 
Notice that the perturbed metric and perturbed scalar field have two sources.
The boundary condition evaluated at the origin leads to
\begin{equation}\label{gravity-boundary}
\left.\partial_z h_{\mu\nu}\right|_0+\left.\frac{U_B}{2}h_{\mu\nu}\right|_0=2\partial_{\mu}\partial_{\nu}\xi_z
-\frac{a^2}{2}\frac{\partial U_B}{\partial \phi}\eta_{\mu\nu}\delta\phi
\end{equation}
The scalar field boundary condition reads
\begin{equation}\label{scalar-boundary}
\left.\partial_z \delta \phi\right|_0=\left.\frac{\partial^2
U_B}{\partial \phi^2}\delta\phi\right|_0
\end{equation}
These equations tell us that the brane fluctuation $\xi_z$ acts as
a source of gravitation on the brane.

Let us now describe the equations of motion in the bulk. 
We first concentrate on the Klein-Gordon equation
\begin{equation}
\Box\delta\phi + \frac{1}{2}h'\phi'  = \frac{1}{2}\frac{\partial^2 U}{\partial
\phi^2} \delta \phi 
\end{equation}
where $h=h^{\mu}_\mu$.
The Klein-Gordon equation admits a solution
\begin{equation}
\delta\phi(x^{\mu},z) = \frac{\partial U_b}{\partial \phi}\hat \phi (x^\mu),
\end{equation}
with
\begin{equation}
h=0
\end{equation}
for any four dimensional massless $\hat \phi (x^{\mu})$. 
It satisfies the boundary condition identically.
Moreover we find that for these solutions the four dimensional metric 
perturbation is traceless. Of course, there might be other solutions 
for which the trace does not vanish; but this solution would not be 
the zero mode, which we consider here.

We are now going to analyse the Einstein equations to first order. 
They are best 
described by using the second order variation of the action.
We consider the coordinate system where the brane is at $z=\xi_z$ and 
therefore write the metric perturbation as $h_{ab}$, where 
$a,b$ is a five--dimensional index and $h_{55}=0=h_{\mu 5}$, as we
consider Gaussian normal coordinates. The second order variation of 
the gravitational Lagrangian reads \cite{anti}
\begin{eqnarray}
\delta^{(2)}S_g &=& \frac{1}{2\kappa_5^2}\int d^5x \sqrt{-g_0} \left(\frac{1}{2}
D_ah^{ab}D_c h^{cb} -\frac{1}{4}D_{a}h_{bc}D^a h^{bc} \right. \nonumber\\
&-&\left. \frac{R}{4}h_{ab}h^{ab}+\frac{1}{2}R_{ab}h^a_ch^{bc}-\frac{1}{2}
R_{abcd}h^{ad}h^{bc}\right )
\end{eqnarray}
while the scalar Lagrangian gives
\begin{equation}
\delta^{(2)}S_{\phi}=-\frac{3}{8\kappa_5^2}\int \sqrt{-g_0}\left(
\frac{{\cal L}_{\phi}}{3}h_{ab}h^{ab}+ (\partial\delta\phi)^2 +
\frac{1}{2}\frac{\partial^2 U}{\partial \phi^2}\delta\phi^2 + 
2h^{ab}\partial_a\delta\phi\partial_b\phi\right)
\end{equation}
where
\begin{equation}
{\cal L}_{\phi}= -\frac{3}{4}\left[ (\partial \phi_0)^2 + U \right]
\end{equation}
Notice that there is a contribution to the gravitational part from the transverse part of the gravitational field.
We can simplify these actions using Einstein's equations
\begin{equation}
R_{ab}=T_{ab}-\frac{1}{3}Tg_{ab}
\end{equation}
and
\begin{equation}
R_{acbd}=-\frac{U_B^2}{16}\left(g^0_{ab}g_{cd}^0-g^0_{ad}g^0_{cb}\right)
+\frac{1}{4}\left(\frac{\partial U_B}{\partial\phi}\right)^2(p_ag^0_{[cd}p_{b]}-
p_cg^0_{a[d}p_{b]})
\end{equation}
where $p_a=\delta_{a5}$. 
The bulk Einstein equations  are then
\begin{equation}\label{bulk-equation}
\Box h_{ab}+\frac{U_B^2}{8}h_{ab} -D_{\{a}D_ch^c_{b\}}= \frac{3}{4}\partial_{\{a}\delta\phi\partial_{b\}}\phi_0
\end{equation}
Notice the unconventional appearance of the transverse part of the metric perturbation.

Let us focus on the $(55)$ component of the Einstein  equations.
It leads to
\begin{equation}
\partial_z \delta\phi=0
\end{equation}
This implies that $\delta{\phi}$ is a function of $x^{\mu}$ only.
Now this is not compatible with the solution of the Klein-Gordon equation unless
\begin{equation}
\delta\phi\equiv 0.
\end{equation}
Therefore the scalar field perturbation vanishes altogether in the
coordinate system where the brane position fluctuates according to
$\xi_z$.

Finally, the ($\mu 5$)--component of the Einstein  equations gives
\begin{equation} 
D_{\nu}h_{~\mu}^{\nu} = 0.
\end{equation}
The metric perturbation is therefore found to be transverse-traceless.
This will be used in the following section to identify the graviton 
wave-function.

In the case where the model with a bulk scalar field is embedded
in supergravity in singular spaces one can understand the absence of an independent scalar degree of freedom as follows. At low
energy the fifth component of the graviphoton completes the
radion field to pertain to a chiral supermultiplet. No
extra scalar degree of freedom is a available to be
paired with the scalar field in order to form the scalar part of
a chiral supermultiplet. This implies that no low energy degree
of freedom can be ascribed to the bulk scalar field.

\subsection{The Graviton}
Now we consider the gravitational equation in order to discuss 
the graviton modes. The field equation for the graviton simplifies to
\begin{equation}
\Box h_{\mu\nu}+\frac{U_B^2}{8}h_{\mu\nu}=0
\end{equation}
together with the boundary condition
\begin{equation}
\left.\partial_z h_{\mu\nu}\right|_0+\left.\frac{U_B}{2}h_{\mu\nu}\right|_0=0
\end{equation}
where we impose $\xi_z=0$. We decompose the solution into 
a product of a $z-$dependent part and a $x^{\mu}-$dependent 
part:
\begin{equation}
h_{\mu\nu} = g(z)\chi_{\mu\nu}(x^\mu).
\end{equation}
The Laplacian operator acting on the gravitational perturbation
can be obtained by writing
\begin{equation}
\Gamma^{\alpha}_{5\beta}=-\frac{U_B}{4}\eta^{\alpha}_{\beta},\
\Gamma^5_{\alpha\beta}=\frac{U_B}{4}g^0_{\alpha\beta},
\end{equation}
where the BPS conditions (\ref{BPS}) have been used.
This leads to
\begin{equation}
\Box h_{\mu\nu}=(\partial_z^2g) \chi_{\mu\nu}+ g\Box\chi_{\mu\nu}
\end{equation}
and eventually
\begin{equation}
\Box \chi_{\mu\nu}=a^{-2}\Box^{(4)}\chi_{\mu\nu}+\left(\left(\frac{\partial
U_B}{\partial \phi}\right)^2 -\frac{3}{8}U_B^2\right)\chi_{\mu\nu}
\end{equation}
We consider the zero modes of the gravitational perturbation
only, i.e. we select
\begin{equation}
\Box^{(4)}\chi_{\mu\nu}=0
\end{equation}
implying that the graviton profile $g$ satisfies
\begin{equation}
\partial_z^2 g+\left(\frac{1}{2}\left(\frac{\partial U_B}{\partial
\phi}\right)^2 - \frac{1}{4}U_B^2\right)g = 0.
\end{equation}
This is a Schr\"{o}dinger equation with a non-trivial potential.
In the Randall-Sundrum case the brane potential is constant
$U_B=4k$ leading to
\begin{equation}
g_{RS}=a^{2}
\end{equation}
which satisfies the boundary condition. 

To find the solution in the general case it is convenient to 
define the conformal coordinate
\begin{equation}
du=\frac{dz}{a};
\end{equation}
with such a choice and putting
\begin{equation}
g= \sqrt a \psi 
\end{equation}
we obtain the graviton equation  
\begin{equation}
-\frac{d^2\psi}{du^2}+ \frac{1}{a^{3/2}}\frac{d^2 a^{3/2}}{du^2}\psi = 0.
\label{gra}
\end{equation}
This is of the form of a supersymmetric quantum mechanics problem with
\begin{equation}
Q^{\dagger}Q\psi=0
\end{equation}
where
\begin{equation}
Q=-\frac{d}{du}+\frac{d\ln a^{3/2}}{du},\ Q^{\dagger}=\frac{d}{du}+\frac{d\ln a^{3/2}}{du}
\end{equation}
The two zero modes are easily found to be
\begin{equation}
\psi_1=a^{3/2}
\end{equation}
and
\begin{equation}
\psi_2= a^{3/2}\int \frac{du}{a^3}.
\end{equation}
The boundary condition can be reexpressed in terms of $\psi$
\begin{equation}
\frac{d(\psi a^{-3/2})}{du}=0
\end{equation}
implying that the graviton profile is given by $\psi_1$, i.e. that
\begin{equation}
g(z)=a^2
\end{equation}
Notice that this is the generalization to models with a bulk scalar field
of the Randall-Sundrum result.
The graviton is then described by
\begin{equation}
h_{\mu\nu}=a^2 \chi_{\mu\nu}
\end{equation}
where $\chi_{\mu\nu}$ is a four-dimensional massless field. 
Now notice that we can use the residual gauge freedom specified by 
$\epsilon_{\mu}(x)$ in order to retrieve the fact that $\chi_{\mu\nu}$      
possesses
two independent polarizations only and therefore qualifies as the
four dimensional graviton. 
This result guarantees that the low energy effective action will
be a scalar--tensor  theory. We are now going to discuss the
appearance of an independent scalar mode whose coupling to gravity 
will be studied. 

\section{The Radion}
So far we have discussed transverse--traceless modes with
$\xi_z=0$. In this section we include fluctuations of the 
brane positions and discuss their dynamics. 
\subsection{Brane Fluctuations}
Coming back to the gravitational equation (\ref{bulk-equation}), 
we can investigate modes with $\xi_z\ne 0$. The boundary condition 
for such modes reads 
\begin{equation}
\left.\partial_z h_{\mu\nu}\right|_0+\left.\frac{U_B}{2}h_{\mu\nu}\right|_0
=2\partial_{\mu}\partial_{\nu}\xi_z
\end{equation}
Taking the trace of the boundary condition and using the
tracelessness of the gravitational perturbation $h_{\mu\nu}$ we find
\begin{equation}
\Box^{(4)}\xi_z\equiv \eta^{\mu\nu}\partial_{\mu}\partial_{\nu}\xi_z=0.
\end{equation}
This equation implies that the brane fluctuation corresponds to a massless
four-dimensional field. It is a mode which complements the graviton
and leads to a scalar-tensor theory at low energy.
One can solve the gravitational equation with the Ansatz
\begin{equation}
h_{\mu\nu}= a^{1/2}\psi_{\xi}(z) \partial_{\mu}\partial_{\nu}\xi_z
\end{equation}
where $\psi_{\xi}$ satisfies (\ref{gra}).
The boundary condition reads
\begin{equation}
\left.\frac{d(\psi_{\xi} a^{-3/2})}{du}\right|_{0}=\left.\frac{2}{a}\right|_{0}
\end{equation}
One can identify the solution with
\begin{equation}
\psi_{\xi}=2a_0^2 \psi_2(z)
\end{equation}
leading to 
\begin{equation}
h_{\mu\nu}=2a_0^2 a^2 \left(\int \frac{dz}{a^4}\right)\partial_{\mu}\partial_{\nu}\xi_z
\end{equation}
It is conspicuous that the two zero modes of the gravitational equation
are transcribed in the two types of physical modes, the graviton and
the brane fluctuation.

\subsection{Including a Second Brane}

So far, we have not used the existence of the second brane. However,
as we shall see in the following, it is important to include a
second brane in this setting. The r{\^o}le of the second brane is
two-fold. On the one hand, in a consistent supersymmetric setting, the 
existence of the second brane is crucial in order to preserve
supersymmetry. On the other hand the second brane screens off the 
naked singularity from the first brane \cite{BD4}.

The metric tensor which is a solution of the gravitational equation can be rewritten
as
\begin{equation}
\hat h_{\mu\nu} = a^2\chi_{\mu\nu} 
+ 2a_0^2 a^2 \left(\int \frac{dz}{a^4}\right)\partial_{\mu}
\partial_{\nu}\xi_z - 2 a^2 \left(\int \frac{dz}{a^2}\right)\partial_{\mu}
\partial_{\nu}\xi_z-\frac{U_B}{2}a^2 \xi_z \eta_{\mu\nu}
\end{equation}
where one recognizes the graviton and the brane fluctuation.
One can define such a solution in the vicinity of each brane.
In the following we shall denote by $a_+$ and $\xi_z^+$ the scale factor
and the brane fluctuation on the positive tension brane 
(respectively $a_-$ and $\xi_z^-$ on the negative tension brane).
In each patch surrounding each brane the metric tensor is single-valued.
We still have to construct a metric tensor which is single-valued 
throughout the bulk.

To do so we first impose that the brane fluctuation solution of (\ref{gra})
is single valued. This is achieved by requiring
\begin{equation}
a_+^2\xi_z^+\equiv a_-^2\xi_z^-\equiv \xi
\end{equation}
where $\xi$ is a four dimensional massless field.
Notice that this mode is intrinsically non-local relating the
fluctuations on one brane to the fluctuations on the other brane. 
The metric tensor is now
\begin{equation}
\hat h^+_{\mu\nu}= a^2\xi_{\mu\nu} + 2 a^2 \left(\int \frac{dz}{a^4}\right)\partial_{\mu}
\partial_{\nu}\xi - 2 \frac{a^2}{a_+^2} \left(\int \frac{dz}{a^2}\right)\partial_{\mu}
\partial_{\nu}\xi-\frac{U_B}{2}\frac{a^2}{a_+^2} \xi \eta_{\mu\nu}
\label{bulk}
\end{equation}
in the patch surrounding the positive tension brane.
In the second patch close to the negative tension brane, the metric tensor is similar
\begin{equation}
\hat h^-_{\mu\nu}= a^2\xi_{\mu\nu} + 2 a^2 \left(\int \frac{dz}{a^4}\right)\partial_{\mu}
\partial_{\nu}\xi - 2 \frac{a^2}{a_-^2}\left(\int \frac{dz}{a^2}\right)\partial_{\mu}
\partial_{\nu}\xi-\frac{U_B}{2}\frac{a^2}{a_-^2} \xi \eta_{\mu\nu}
\end{equation}
Obviously the metric tensor cannot be patched up in this form as
the two expressions in the two patches are different.

However one can perform a change of coordinates which preserves the 
Gaussian normal coordinates and translates the second brane. In the 
second patch it reads
\begin{equation}
z\to z - \left(\frac{1}{a_-^2}-\frac{1}{a_+^2}\right)\xi
\label{change}
\end{equation}
which transforms $h_{\mu\nu}^-$ into $h_{\mu\nu}^+$.
The negative tension brane sits now at
\begin{equation}
T(x)= r_c - \left(\frac{1}{a_-^2}-\frac{1}{a_+^2}\right)\xi.
\end{equation}
Now the bulk metric is single-valued and given by
(\ref {bulk}).
Notice that the position of the second brane fluctuates according to fluctuations of the 
massless field $\xi$.
Moreover the field $T(x)$ can now be interpreted as the radion measuring the
size of the extra dimension. The radion is also a four dimensional massless
field. In particular the radion measures the brane bending of the
second brane. 

We would like to obtain the metric tensor in a form where the
size of the interval does not fluctuate, i.e. the
radion would appear as a fluctuation of the metric.
This can be achieved thanks to  a new change of coordinates defined
by
\begin{equation}
z\to z + \left(\frac{1}{a^2}-\frac{1}{a_+^2}\right)\xi
\label{che}
\end{equation}
which rescales the size of the bulk in such a way that the second brane
sits at $z=r_c$.
Imposing that no $(\mu z)$ component appears in the metric tensor implies
that one must also transform
\begin{equation}
x^{\mu}\to x^{\mu} + \xi^{\mu}
\end{equation}
where
\begin{equation}
\xi_{\mu}= -\frac{a^2}{a_+^2}\int dz \left(\frac{a_+^2}{a^4} - 
\frac{1}{a^2}\right)\partial_{\mu}\xi
\end{equation}
This implies that the metric tensor becomes
\begin{equation}
\hat h_{\mu\nu}= a^2\chi_{\mu\nu}-\frac{U_B}{2}\frac{a^2}{a_+^2}\xi \eta_{\mu\nu}
\end{equation}
Let us now write the full metric including both the background
and the perturbation
\begin{eqnarray}
ds^2 &=& a^2 \left( z +\left(\frac{1}{a^2}
-\frac{1}{a_+^2}\right)\xi\right)dx^{\mu}dx_{\mu} \nonumber\\
&+&\left(1+\partial_z\left(\frac{1}{a^2}-\frac{1}{a_+^2}\right)\xi\right)^2dz^2+
\left(a^2\chi_{\mu\nu}-\frac{U_B}{2}\frac{a^2}{a_+^2}\xi \eta_{\mu\nu}\right)dx^{\mu}dx^{\nu}
\end{eqnarray}
This can be simplified to
\begin{equation}
ds^2=a^2(G(z,x))g_{\mu\nu}dx^{\mu}dx_{\nu} + (\partial_z G)^2 dz^2 
\end{equation}
to linear order. We have defined 
\begin{equation}
G(z,x)=z+\frac{\xi}{a^2}, g_{\mu\nu}=\eta_{\mu\nu}+\chi_{\mu\nu}
\end{equation}
This is the generalization of the metric Ansatz proposed by
Charmousis, Gregory and Rubakov \cite{Charmousis}.

Let us now derive the fluctuation of the scalar field in the various
coordinate systems.
When the two branes are  translated we have $\delta\phi=0$.
Shifting the first brane to the origin implies that
\begin{equation}
\delta\phi^+=\frac{U_B'}{a_+^2}\xi
\end{equation}
in the first patch and
\begin{equation}
\delta\phi^-=\frac{U_B'}{a_-^2}\xi
\end{equation}
in the second patch.
Let us now perform the local change of coordinates in the second patch
(\ref{change}). This defines a single-valued scalar field fluctuation
which coincides with $\delta \phi^+$ throughout the bulk.
The change of coordinates (\ref{che})
leads to a single-valued fluctuation
\begin{equation}
\delta\phi=U_B'\frac{\xi}{a^2}
\end{equation}
Combined with the background scalar field this leads to
\begin{equation}
\phi(z,x)=\phi_0(G(z,x))
\end{equation}
Notice that the scalar field is now a function of the brane
fluctuation parameter $\xi$, i.e. the radion. There is no
scalar field dynamics {\it per se}.

\subsection{The Scalar-Tensor Theory at Low Energy}

Now that we have described the independent low energy degrees of freedom
of the theory with a bulk scalar field we are in a position to
discuss the low energy scalar tensor theory.  We will only examine two relevant terms,
i.e. the Einstein-Hilbert term and the radion kinetic term.

The Einstein-Hilbert kinetic term can be seen to be
\begin{equation}
\int d^4 x \sqrt {-g} \Phi R
\end{equation}
where $g_{\mu\nu}$ is the four dimensional metric and the effective
Newton constant is
\begin{equation}
\Phi= \frac{1}{\kappa_5^2}\int_{\xi/a_{+}^2}^{r_c+\xi/a_-^2}a^2(y) dy
\end{equation}
Notice the explicit dependence on the radion field.
Similarly the kinetic terms for the radion field is
\begin{equation}
-\frac{3}{8\kappa^2_5} \int d^4 x \sqrt{-g} K \left(\partial \xi\right)^2,
\end{equation}
where 
\begin{equation}
K = \int_0^{r_c} dz a^{-2}\left[U_B^2 
+ 2 \left(\frac{\partial U_B}{\partial \phi}\right)^2 \right].
\end{equation}
Notice that this is always positive.

Let us focus on the $AdS_5$ case where the scalar tensor theory is specified
by
\begin{equation}
\Phi= \frac{1}{2k\kappa_5^2}(e^{-k\xi/a_+^2}-e^{-kr_c-k\xi/a_-^2})
\end{equation}
and
\begin{equation}
K=8k(e^{2kr_c}-1)
\end{equation}
In the large $r_c$ limit and for small fluctuations of the radion
$\xi$ this result coincides with the results of Garriga and Tanaka
\cite{Garriga} on the Brans-Dicke parameter when coupling the theory
to matter.  For the general case the discussion of the coupling to
matter for brane world models with a bulk scalar is left for a
companion paper.
\section{Conclusions}
In this paper we have given a detailed discussion about the dynamics of 
a two--brane system with bulk scalar field. We have investigated 
the coupling of the distance between the branes, which is a 
scalar degree of freedom (radion) and  four--dimensional gravity. When the BPS condition (\ref{BPS}) is imposed on the 
system, we find that the bulk scalar field and the radion combine to give
one independent scalar degree of freedom; i.e.  the bulk scalar field does 
not give rise to an additional degree of freedom. 
At low energy we find that the theory reduces to a scalar--tensor theory coupling the radion to gravity. 

There is one final step to be taken, namely the breaking of supersymmetry on
the brane, before we are in a position to confront the brane world with a
bulk scalar field with observation. It is likely that breaking supersymmetry
will induce a potential term for the radion field in the low energy effective
action. This is in progress. 

\vspace{0.5cm}\noindent{\bf Acknowledgements:} We are grateful for 
discussions with C. Charmousis, R. Gregory, L. Pilo and T. Wiseman. This work 
was 
supported by PPARC (A.C.D. \& C.S.R.), the Deutsche Forschungsgemeinschaft 
(C.v.d.B.), a CNRS--Royal Society exchange grant for collaborative research
and the European network (RTN), HPRN-CT-2000-00148 and 
PRN-CT-2000-00148.

\end{document}